\documentclass[a4paper,11pt]{article}
\pdfoutput=1 
\usepackage{color}
\usepackage{jcappub} 

\usepackage[caption=false]{subfig}
\usepackage{graphicx}
\usepackage{bm}
\usepackage{dsfont}
\usepackage{epsf}
\usepackage{braket}
\usepackage{rotating}
\usepackage{epsfig,graphics,rotate,color}
\usepackage{wrapfig}
\usepackage{amssymb}
\usepackage{amsmath}
\usepackage{amsfonts}
\usepackage[utf8]{inputenc} 
\usepackage{booktabs}
\usepackage{gensymb}
\usepackage{placeins}
\usepackage{diagbox}
\usepackage[dvipsnames]{xcolor}
\usepackage{tabularx}
\usepackage{soul} 
\usepackage[normalem]{ulem} 

\def\al{\alpha}
\def\be{\beta}
\def\ga{\gamma}
\def\de{\delta}
\def\ep{\epsilon}

\def\th{\theta}

\def\si{\sigma}

\def\ta{\tau}

\def\ph{\phi}

\def\nue{\nu_e}
\def\numu{\nu_\mu}
\def\nutau{\nu_\tau}

\def\lsim{\mathrel{\rlap{\lower4pt\hbox{\hskip1pt$\sim$}}
    \raise1pt\hbox{$<$}}}
\def\gsim{\mathrel{\rlap{\lower4pt\hbox{\hskip1pt$\sim$}}
    \raise1pt\hbox{$>$}}}

\newcommand{\beq}{\begin{eqnarray}}
\newcommand{\eeq}{\end{eqnarray}}

\def\ofratio{(\phi_e^\oplus:\phi_\mu^\oplus:\phi_\tau^\oplus)}

\def\ofratios{(\phi_e^\oplus:\phi_\mu^\oplus:\phi_\tau^\oplus:\phi_s^\oplus)}
\def\ifratios{(\phi_e^p:\phi_\mu^p:\phi_\tau^p:\phi_s^p)}

\title{
Sterile Neutrinos in Astrophysical Neutrino Flavor
}

\abstract{In this paper, we study the effect of active-neutrino-sterile-neutrino mixing in the expected high-energy astrophysical neutrino flavor content. Non-unitarity in the measurement of the three-active neutrinos can be due to the existence of sterile neutrino states. We introduce the concept of the four-flavor tetrahedron in order to visualize the lack of unitarity in the astrophysical neutrino three-flavor triangle. We demonstrate that active-sterile neutrino mixings modify the allowed region of the astrophysical flavor ratio from the standard case. However, a projection of the four-flavor tetrahedron has restrictions of phase space similar to the three-flavor triangle. On the other hand, the initial presence of astrophysical sterile neutrinos drastically changes the scenario, and it allows an apparent unitarity violation in the three-flavor triangle space. Using current global fit constraints  including the non-unitarity case, we also illustrate the allowed astrophysical neutrino flavor ratios. Thus, the measurement of the high-energy astrophyscal neutrino flavor content allows us to explore sterile neutrinos independently of the sterile neutrino mass scale. These are topics of investigation for current and future neutrino telescopes.}

\author[a]{Carlos A.~Arg\"uelles}
\emailAdd{caad@mit.edu}
\author[b,c]{Kareem~Farrag}
\emailAdd{k.r.h.farrag@qmul.ac.uk}
\author[d]{Teppei Katori}
\emailAdd{teppei.katori@kcl.ac.uk}
\author[e]{Rishabh Khandelwal}
\emailAdd{rkhandelwal3@wisc.edu}
\author[b]{Shivesh Mandalia}
\emailAdd{s.p.mandalia@qmul.ac.uk}
\author[f]{Jordi Salvado}
\emailAdd{jsalvado@icc.ub.edu}

\affiliation[a]{Department of Physics, Massachusetts Institute of Technology, Cambridge, MA 02139, USA}
\affiliation[b]{School of Physics and Astronomy, Queen Mary University of London, London E1 4NS, UK}
\affiliation[c]{Physics and Astronomy, University of Southampton,
Southampton, SO17 1BJ, U.K}
\affiliation[d]{Department of Physics, King's College London, London WC2R 2LS, UK}
\affiliation[e]{Department of Physics and Wisconsin IceCube Particle Astrophysics Center, University of Wisconsin, Madison, WI 53706, USA}
\affiliation[f]{Departament de F\'isica Qu\`antica i Astrofísica and Institut de Ciencies del Cosmos, Universitat de Barcelona, Diagonal 647, E-08028 Barcelona, Spain}

\notoc 
\begin{document} 
\maketitle
\flushbottom
\date{\today}

\section{Introduction}
\label{sec:intro}

High-energy neutrino astronomy uses neutrinos that have the longest baselines and the highest energies to inform us of astrophysical environments. In these extreme regimes, the fundamental properties of neutrinos are poorly constrained. Thus, it is timely to examine how the uncertainties in neutrino properties impact our understanding of neutrino sources while simultaneously introducing new ways of performing tests of new physics given astrophysical neutrino data~\cite{Arguelles:2019rbn}. High-energy astrophysical neutrinos, first observed by the IceCube Neutrino Observatory~\cite{Aartsen:2013jdh,Aartsen:2014gkd}, offer new probes of the Standard Model (SM) paradigm that includes three active massive neutrinos, the so-called neutrino Standard Model ($\nu$SM).

Experimentally it is verified that the neutrino flavor eigenstates, produced in charged-current electroweak interactions, are a linear superposition of neutrino mass eigenstates. These eigenstates are related by a matrix known as the lepton mixing matrix. In the $\nu$SM, the lepton mixing matrix is unitary and coincides with the Pontecorvo–Maki–Nakagawa–Sakata (PMNS) matrix. The parameters that determine the mixing matrix, which we denote as $\mathbf{U}$, have been studied in numerous Earth-based experiments over a range of energies of $O(\text{MeV}-\text{GeV})$. In the case of Dirac neutrinos, this matrix can be parameterized by means of three mixing angles and one CP-violating phase. If neutrinos are Majorana, two additional phases exist, but their effects cannot be probed by studying neutrino flavor morphing. These angles have been measured to near percent-level precision by a combination of solar, reactor, atmospheric, and accelerator neutrino experiments~\cite{Esteban:2018azc}. Current long-baseline accelerator-sourced oscillation experiments, such as T2K~\cite{Abe:2018wpn} and NOvA~\cite{NOvA:2018gge}, have shown preference for nonzero Dirac-CP violation, ruling out a null phase paradigm at over 95\% C.L. Furthermore, global fits to the current data favor normal neutrino mass ordering~\cite{Esteban:2018azc}. Future oscillation experiments, such as DUNE~\cite{Acciarri:2016crz}, Hyper-Kamiokande~\cite{Abe:2015zbg}, JUNO~\cite{An:2015jdp}, RENO-50~\cite{Seo:2015yqp}, ORCA~\cite{Adrian-Martinez:2016fdl}, and the IceCube Upgrade~\cite{Ackermann:2017pja} will further reinforce these results. 

Due to theoretical motivations, the leptonic mixing matrix is assumed to be unitary in the $\nu$SM. Within this paradigm, we often express the global neutrino data in terms of three mixing angles and one phase. In the NuFit~4.0 global fit to neutrino oscillation data\footnote{The fit does not include Super-Kamiokande atmospheric neutrino data, which is the default of NuFit~4.0.}~\cite{Esteban:2018azc}, these mixing angles were found to be: $\th_{12}=33.82_{-2.21}^{+2.45}$, $\th_{23}=49.6_{-9.3}^{+2.8}$, $\th_{13}=8.61_{-0.39}^{+0.38}$, and $\de_{CP}=215_{-90}^{+177}$ at the $3\sigma$ C.L. It is then possible to translate the uncertainties on these angles into uncertainties on the PMNS matrix elements. Doing this results in
\beq
  |U|_{3\si}^{PMNS} =
 \left(
\begin{matrix}
|U_{e 1}|  & |U_{e 2}| & |U_{e 3}|  \\
|U_{\mu 1}|& |U_{\mu 2}|& |U_{\mu 3}|\\
|U_{\ta 1}|& |U_{\ta 2}|& |U_{\ta 3}|
\end{matrix}
\right)
=
\left(
\begin{matrix}
0.797\rightarrow 0.842 & 0.518\rightarrow 0.585 & 0.143\rightarrow 0.156 \\
0.233\rightarrow 0.495 & 0.448 \rightarrow 0.679 & 0.639 \rightarrow 0.783\\
0.287 \rightarrow 0.532 & 0.486 \rightarrow 0.706 & 0.604\rightarrow 0.754
\end{matrix}
\right).
\label{eq:unitarity}
\eeq

However, the PMNS matrix elements can be found individually without assuming the twelve unitarity conditions. Once such constraints are removed, the allowed elements can look very different~\cite{Parke:2015goa}, as shown below:

\beq
  |U|_{3\si}^{w\/o~unitarity}
=
\left(
\begin{matrix}
0.76\rightarrow 0.85 & 0.50\rightarrow 0.60 & 0.13\rightarrow 0.16\\
0.21\rightarrow 0.54 & 0.42\rightarrow 0.70 & 0.61\rightarrow 0.79\\
0.18\rightarrow 0.58 & 0.38\rightarrow 0.72 & 0.40\rightarrow 0.78
\end{matrix}
\right).
\label{eq:nonunitarity}
\eeq

Note that the tests of the unitarity conditions are much weaker when compared with the ones performed on the Cabibbo-Kobayashi-Maskawa (CKM) matrix~\cite{Tanabashi:2018oca}, the quark sector analogue to the PMNS matrix. At present, the PMNS matrix can allow for large non-unitarity, where the deviation of normalizations from unity can reach 40\%~\cite{Parke:2015goa} at $3\sigma$ C.L. 

It is interesting to pursue this because there are persistent anomalies in many oscillation experiments. These arise in short-baseline neutrino experiments, collectively coined as the short-baseline anomalies  ~\cite{Aguilar:2001ty,Giunti:2010zu,Mention:2011rk,Aguilar-Arevalo:2018gpe}; for a recent review see~\cite{Giunti:2019aiy,Diaz:2019fwt,Boser:2019rta}. They can be explained by adding an additional neutrino mass state with a mass-square difference of around $\mathcal{O}(1 \text{ eV}^2)$. This additional neutrino mass state does not participate in the electroweak interactions and is thus known as a sterile neutrino. The amount of non-unitarity that is allowed in the PMNS matrix is comparable and compatible with results from global fits to eV-scale sterile neutrinos~\cite{Collin:2016aqd}. However, an eV-scale sterile neutrino interpretation of the LSND-MiniBooNE signal, now with a combined significance of more than $5\si$, is in strong tension with other experiments~\cite{Dentler:2018sju}; in particular when compared to the lack of muon-neutrino disappearance in MINOS~\cite{Adamson:2017uda}, Daya Bay~\cite{Adamson:2016jku}, and IceCube~\cite{TheIceCube:2016oqi,Aartsen:2017bap}. This has prompted the community to look for alternative explanations to the MiniBooNE-observed data excess.  These include SM neutral current photon production~\cite{Hill:2010zy,Zhang:2012xn,Wang:2014nat,Kullenberg:2011rd,Abe:2019cer} or beyond the SM (BSM) physics (for example, $Z'$ production~\cite{Jordan:2018qiy,Ballett:2016opr,Bertuzzo:2018itn,Arguelles:2018mtc,Bustamante:2015waa, Rasmussen:2017ert}). If the short-baseline anomaly is confirmed to be due to sterile neutrinos, the PMNS matrix would be required to span $3+N$ dimensions, with $N$ at least one. In fact, the presence of sterile neutrinos, or SM singlet states, is natural for any masses other than the eV-scale~\cite{DiBari:2016guw}. In this situation, it is desired to look for sterile neutrinos for all possible mass ranges. If such a state exists, we would see that the PMNS matrix is non-unitary because of the leakage of the probability to sterile states~\cite{Fong:2017gke}.

A relatively new approach involves using high-energy astrophysical neutrinos to search for signatures of non-unitarity. One useful observable in this endeavor is the astrophysical neutrino flavor composition. IceCube has measured these neutrinos' flavor composition and reported it in terms of the flavor ratio, since this observable is weakly dependent on the poorly constrained flux normalization~\cite{Aartsen:2015ivb,Aartsen:2015knd,Aartsen:2018vez,Palomares-Ruiz:2014zra,Palladino:2015zua}. It is commonly displayed using the flavor triangle to represent the flavor ratio of astrophysical neutrinos~\cite{Fogli:1995uu,Athar:2000yw,gaisser2016cosmic}. This flavor information is interesting because the presence of sterile neutrino states causes non-unitarity in the PMNS matrix, which may thus show up in terms of anomalous flavor structure on Earth. Searching for new physics using flavor ratios has been discussed by many authors; see for instance~\cite{Arguelles:2015dca,Bustamante:2015waa,Bustamante:2016ciw,Bustamante:2018mzu,Rasmussen:2017ert,Brdar:2016thq,Klop:2017dim,Farzan:2018pnk} and references therein. If the allowed active-sterile mixing has a very small mass splitting, non-unitarity of the PMNS matrix can show up only for neutrinos that undergo cosmological-scale propagation~\cite{Learned:1994wg}, such as astrophysical neutrinos. Sterile neutrinos of extremely small mass splittings have not been tested, and they could have a significant contribution to astrophysical neutrino flavor non-unitarity.

In this paper, we relax the assumption that the neutral-lepton mixing matrix is unitary and study the expected regions of astrophysical flavors at Earth that are attainable given the standard scenarios of astrophysical neutrino production. We make a series of predictions of the expected flavor ratios at Earth by extending the three-flavor triangle to a four-flavor tetrahedron. Finally, we show the predictions for the astrophysical neutrino flavor ratio including current knowledge of non-unitarity from global oscillation data. The outline of the remainder of this paper is as follows: in Section~\ref{sec:theory} we discuss the theoretical background pertaining to unitarity and the modification non-unitarity provides to neutrino oscillations when extra mass eigenstates are present. We introduce a direct measure of non-unitarity, a parameter $\epsilon$, which measures the amount of non-unitarity. We also introduce the $SU(4)$ Haar measure and the four-flavor tetrahedron, which we will use to represent the set of four numbers that add to a constant to map out the full phase space of the active-neutrino-sterile-neutrino mixing. In Section~\ref{sec:mixing1}, we study sterile neutrino mixing without the presence of a sterile neutrino state at the source. In Section~\ref{sec:mixing2}, we study sterile neutrino mixing; however, this time we assume the presence of a sterile neutrino state at the source. In Section~\ref{sec:mixing3}, we discuss how these effects can be probed with current and next-generation neutrino telescopes. Finally, we make our conclusions and elucidate future capabilities and studies necessary to bolster our new physics search through the study of high-energy astrophysical neutrino flavor composition.

\section{Theory and Methodology}
\label{sec:theory}

The astrophysical neutrino flavor composition is a powerful observable for studying both astrophysical neutrino production mechanisms and oscillation parameters~\cite{Pakvasa:2007dc}. The current uncertainty in this observable does not allow us to differentiate between astrophysical neutrino flavor compositions~\cite{Aartsen:2015ivb,Aartsen:2015knd,Aartsen:2018vez,Palladino:2015zua} at their source. However, it has been pointed out that the astrophysical neutrino flavor composition can explore new physics despite this situation, because the unitary evolution of the flavor structure is predictable and the flavor ratio on Earth can take only certain values even with new physics~\cite{Arguelles:2015dca,Ahlers:2018yom}. To predict the flavor content at Earth, two ingredients are needed: an assumption about the measure of the neutral lepton mixing matrix and an assumption on the source flavor composition. For the former, we choose uniform sampling in the $SU(3)$ space under the Haar measure, or so-called ``anarchy'' sampling~\cite{Haba:2000be}.

The $SU(3)$ Haar measure describes the unit volume of the leptonic mixing matrix and can be fully described using the mixing angles and phase. Omitting the Majorana phases, as they do not affect neutrino oscillations, this measure can be written as
\beq
d U = d s^2_{12} \wedge d c^4_{13} \wedge d s^2_{23} \wedge d\delta~,
\eeq
where $s_{ij}$, $ c_{ij}$, and $\delta$ correspond to sine, cosine, and phase of the angle that parameterize the PMNS matrix according to the parameterization given in~\cite{Tanabashi:2018oca}. Using this measure and assuming three-neutrino unitarity, the flavor ratio probability density on Earth can be computed for each initial flavor ratio at production. For example, a flavor ratio at Earth of $(1:1:1)$ is statistically more likely due to the larger phase-space density when we sample uniformly over the Haar measure. This results in initial admixtures ending up closer to the center of the flavor triangle due to mixing~\cite{Arguelles:2015dca}. In fact, this feature motivates the proposed anarchic structure in the PMNS matrix~\cite{Haba:2000be}. 

A variety of new physics scenarios are proposed within three-flavor unitary mixing~\cite{Arguelles:2015dca,Bustamante:2016ciw,Bustamante:2018mzu,Rasmussen:2017ert}. In this work, we will extend the treatment of unitary evolution beyond the three flavors of active neutrinos to account for possible non-unitary behavior in the neutral lepton mixing matrix~\cite{Parke:2015goa}. In order to discuss the extension of oscillation probabilities with more neutrino species, we follow the notation introduced in~\cite{Fong:2017gke,Blennow:2016jkn}. We can extend the known neutrino system to $3+N$ flavor states and $3+M$ mass states by writing the matrix that relates neutrinos mass and flavor states in vacuum as 
\beq
\mathbf { U } = \left[ \begin{array} { l l } { U } & { W } \\ { Z } & { V } \end{array} \right],
\label{eq:extended_pmns}
\eeq
where $U$ ($3\times 3$) is the matrix that relates the active flavors to the first three neutrino mass states, $W$ ($3 \times M$) is the matrix that encodes the mixing between active flavors and additional mass states, $Z$ ($N \times 3$) is the matrix that shows the relationship between sterile flavors and the first three mass states, and $V$ ($N \times M$) is the matrix between the sterile flavor states and the subsequent mass states.

We need to fix the dimension of the space $N$ so that we can uniformly select matrices respecting the Haar measure in the given $SU(3+N)$ space~\cite{Brdar:2016thq}. Because of this, we assume $M=N$, {\it i.e.} flavor and mass states always have the same dimension. Furthermore, we assume $N=1$ so that we sample uniformly over the set of unitarity $4\times 4$ matrices according to the Haar measure for the locally compact group $SU(4)$. In fact, experimental measurements only under-constrain mixing matrix elements with dimensions greater than three, and $N=1$ is the most optimal. We believe the results at higher dimensions, $N=2, 3...$, simply extend from our investigations in this paper. This choice fixes the dimension of the mixing matrix, and the volume element of the neutrino mixing is described by the Haar measure of $SU(4)$~\cite{Brdar:2016thq},

\beq
d\mathbf{ U} = d s^2_{12}\wedge dc^4_{13}\wedge d s^2_{23}\wedge dc^6_{14}\wedge dc^4_{24}\wedge ds^2_{34}\wedge d\delta \wedge d\delta_1 \wedge d\delta_2.
\eeq

In order to describe the phase space density in the flavor tetrahedron, we sample the mixing matrix from this space.

Under the situation we consider, neutrinos lose coherence completely after the long propagation ($\gg 1$Mpc), and they do not oscillate but mix. The probability of such neutrino mixing or time-averaged oscillations can be written using only the mixing matrix elements: 

\begin{equation}
  P_{\al\be} = \sum_{i=1}^{4}|\mathbf{U}_{\alpha i}|^2 |\mathbf{U}_{\beta i}|^2.
  \label{eq:prob}
\end{equation}
        
Here, $\mathbf{U}_{\alpha i}$ is sampled from the aforementioned $SU(4)$ Haar measure. The greek index $\alpha$ runs over $e$, $\mu$, $\ta$,~and $s$, and $i$ indicates mass state. It is unitary in four-dimensional space, but it is non-unitary in $U$ ($3\times 3$). Non-unitarity neutrino oscillations have been studied in the past~\cite{Fong:2017gke,Blennow:2016jkn}, but these studies' main focus was terrestrial oscillation experiments. In such cases, people usually study oscillation probabilities between given flavors. This is not the case for the astrophysical neutrino flavor physics where initial flavors are not known and we can only assume their ratios. Furthermore, we assume all flavors have the same spectral index within a given production model. In this way, the analysis is independent from the absolute flux normalization and spectral index, which are actively studied but not well known ~\cite{Aartsen:2015rwa}. Thus, in this paper, the flux of astrophysical neutrino flavor $\be$ is denoted by its relative normalization, $\ph_\be$. We use $\al_\be$ for the flavor fraction normalized to $1$, namely $\al_\be=\ph_\be/\sum_\ga \ph_\ga$.  Neutrino telescopes are insensitive to the signs of measured charged lepton. This means we define each $\ph_\be$ as a sum of both neutrino and antineutrino states. We also define a superscript index $p$ to represent the flavor at the production site, and $\oplus$ at the detection. Thus, the flavor fraction of astrophysical neutrino flavor $\be$ at Earth can be written in the following way, 

\begin{equation}
\phi_{\be}^{\oplus} = \sum_{\al}P_{\al\be}\,\phi_{\al}^p.
\label{eq:earthratio}
\end{equation}

Here, $P_{\al\be}$ is defined in Eq.~\eqref{eq:prob}. It is natural in the case of three neutrino flavors $\ofratio$ to plot the distribution of possible ratios on a flavor triangle as a two dimensional projection, since the normalized flavor ratio satisfies the constraint $\al^p_e+\al^p_\mu+\al^p_\ta=\al^{\oplus}_e+\al^{\oplus}_\mu+\al^{\oplus}_\ta=1$. We generalize this method in order to consider the $(3+1)$ case with $\ofratios$, having introduced a new sterile flavor component, $\ph_s$. In illustrating the parameter space, we assume unitarity in the four-dimensional leptonic mixing matrix, $\al^p_e+\al^p_\mu+\al^p_\ta+\al^p_s=\al^{\oplus}_e+\al^{\oplus}_\mu+\al^{\oplus}_\ta+\al^{\oplus}_s=1$.

\begin{figure}
\centering
\includegraphics[scale=0.3]{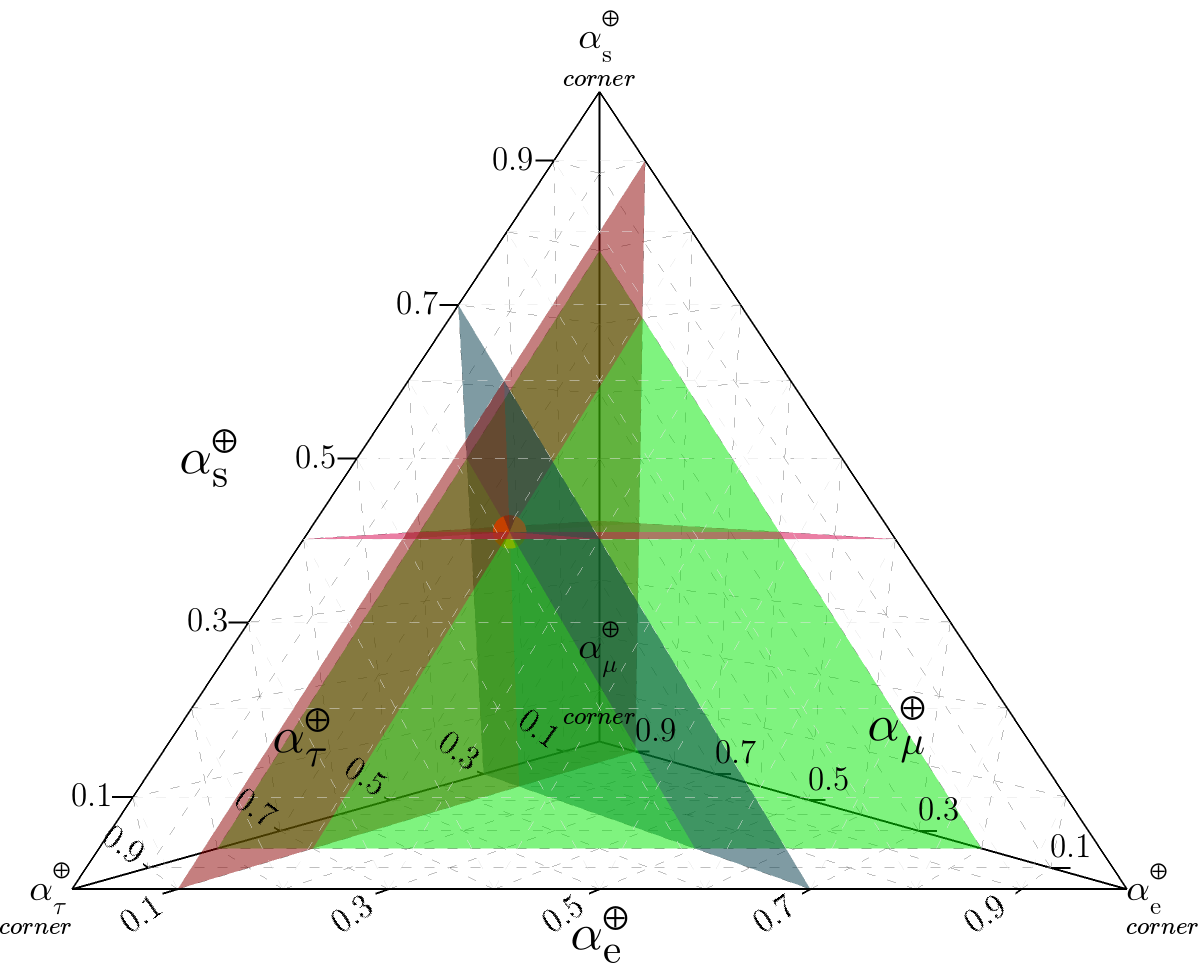}
\caption{An example to demonstrate how to read the four-flavor tetrahedron diagram. The point in orange corresponds to flavor ratio $(0.1:0.2:0.3:0.4)$. We read the flavor tetrahedron points as the altitude from each corner, respectively. This is shown using four planes in the figure. In this case, the red plane opposite the $\alpha_e^{\oplus}$ corner sits at  $\alpha_e^{\oplus}=0.1$, the green plane opposite the $\alpha_\mu^{\oplus}$ corner (hidden behind the tetrahedron) at $\alpha_\mu^{\oplus}=0.2$, the blue plane opposite $\alpha_\tau^{\oplus}$ lies at $\alpha_\tau^{\oplus}=0.3$, and the purple horizontal plane opposite the $\alpha_s^{\oplus}$ corner at $\alpha_s^{\oplus}=0.4$.}
\label{fig:aid}
\end{figure}

There are three astrophysical production mechanisms that we focus on. First, we have the production of neutrinos through pion decay, with source ratio $(1:2:0:x)$. Here we leave the sterile component $x$ free to float, as in principle sterile neutrinos could be produced at the source  with the active neutrinos. The second mechanism is neutron decay, which emits neutrinos in the flavor ratio $(1:0:0:x)$. The final case is muon damping, whereby incoming muons do not produce high-energy neutrinos, producing a flavor ratio $(0:1:0:x)$. Figure~\ref{fig:aid} shows an example of a four-flavor tetrahedron diagram. Note, we do not consider the tau neutrino dominant scenario $(0:0:1:x)$ but it is possible under certain new physics models, such as dark matter decay~\cite{DiBari:2016guw}. The flavor tetrahedron represents a higher dimensional analogy to the flavor triangle. Here we define a convention for the visualization of a point in the tetrahedron such that the labels are consistent with those in the three flavor case for the bottom triangle. In the flavor tetrahedron, for any given point with coordinates $(\al^{\oplus}_e,\al^{\oplus}_\mu,\al^{\oplus}_\ta,\al^{\oplus}_s)$, the geometrical altitudes are equal to $1-\al^{\oplus}_e$, $1-\al^{\oplus}_\mu$, $1-\al^{\oplus}_\ta$, $1-\al^{\oplus}_s$ for each respective component. These are equal to the distance perpendicular to the base opposite the corner of maximal flavor, projected onto the corresponding flavor axis, as shown in Figure~\ref{fig:aid}.

Now, we introduce our measure of non-unitarity. The parameter $\ep$ denotes a sufficient measure of non-unitarity, given by the norm of the $3\times 3$ lepton mixing submatrix $U$ and containing only contributions from the $W$ component of $\mathbf{U}$. We define this as

\begin{equation}
\epsilon = ||U\cdot U^\dagger - \mathds{1}||_2 = ||W\cdot W^\dagger||_2~.
\end{equation}

Here, $||\cdot||_2$ is the matrix 2-norm, or so-called spectral norm. By definition, the 2-norm describes the magnitude of the matrix $U\cdot U^{\dagger}-\mathds{1}$, and in our case, it is sufficient to define how far $U$ is from being unitary. Then there are two distinct effects we can probe using this method: how changing the amount of non-unitarity $\ep$ affects the available parameter space and how modifying the sterile component $x$ at the source alters the available flavor ratios for our randomly sampled mixing matrices.

\subsection{Algorithms for generating matrices}
This section outlines the technique we use to generate random unitary matrices. An algorithm can be found in the appendix. As previously mentioned, Majorana phases do not influence neutrino oscillations, so note $U(4)=SU(4)\times U(1)$ or $SU(4)$ will produces equal distributions. In this work, we follow the technique of Gram-Schmidt orthonormalization to generate a $4\times 4$ unitary matrix (see \cite{mezzadri2006generate,eaton1983multivariate} and references therein). After we generate a unitary matrix, we calculate the $\epsilon$ for each matrix, and collect the matrices with the same $\epsilon$ into sets. We then compute the probability associated to each matrix, and given some source flavor ratio assumed, compute the expected flavor ratio at the Earth. 

\begin{figure}
\centering
\begin{tabular}{cc}
\includegraphics[width=0.43\textwidth,trim={0 0 5cm 0}, clip]{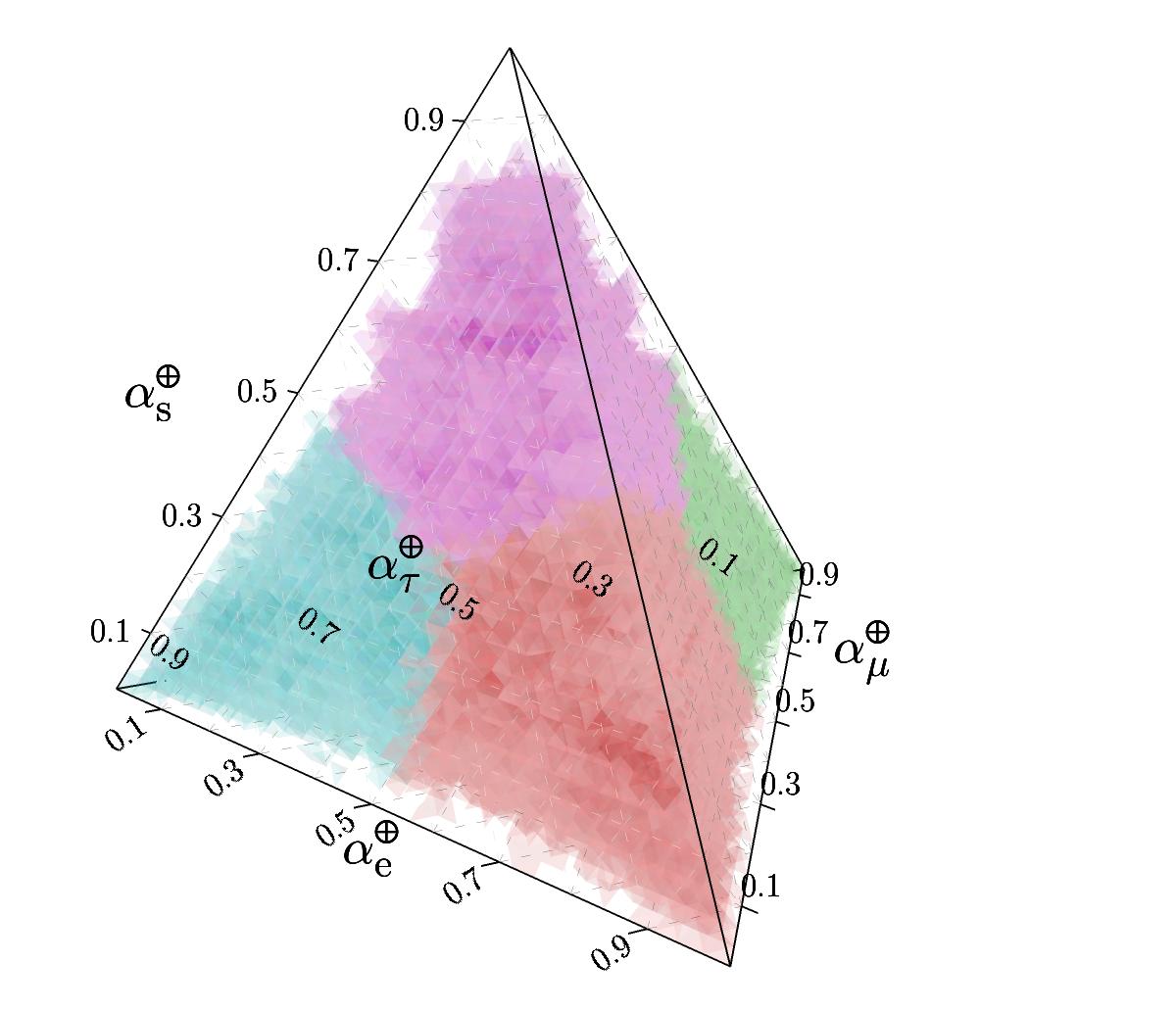} 
 & \includegraphics[width=0.43\textwidth]{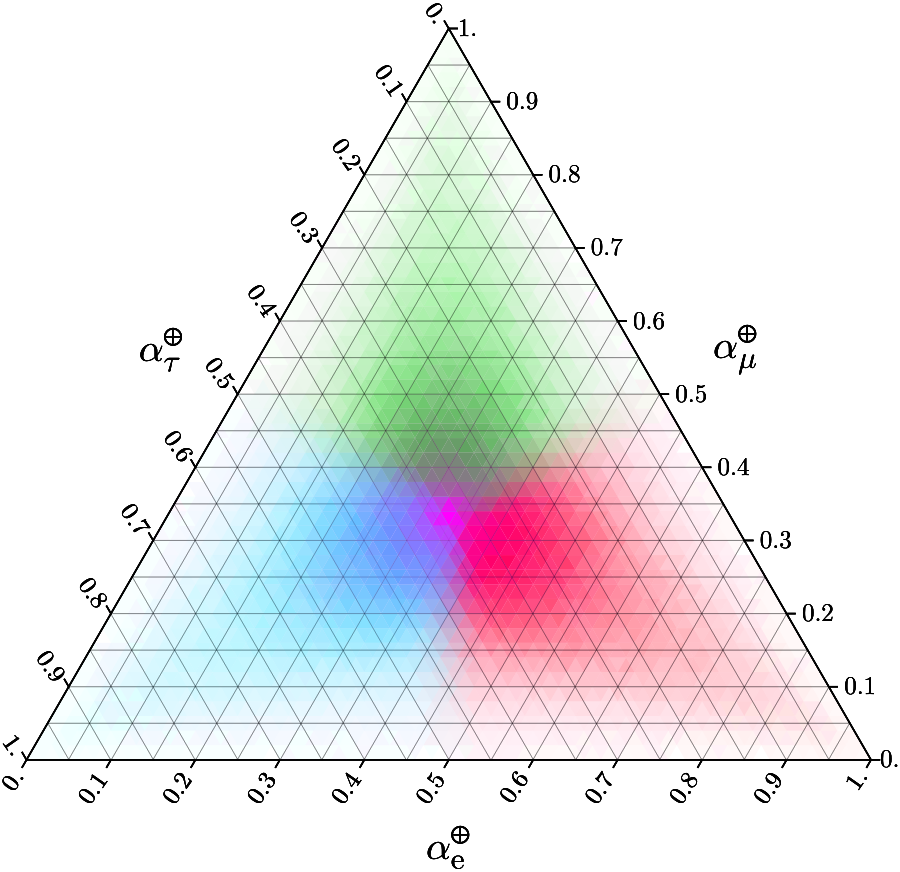}
\end{tabular}
\caption{Left figure shows the full phase space generated $SU(4)$ Haar measure with four pure flavor states at the source. Here, the red volume is made from $\nue$ only $(1:0:0:0)$, the green volume is for $\numu$ only $(0:1:0:0)$, the blue volume is for $\nutau$ only $(0:0:1:0)$, and the purple volume is for $\nu_s$ only $(0:0:0:1)$. Right figure shows the projection of the four-flavor tetrahedron space onto the three-flavor triangle space. Note that in the case of nonzero $\nu_s$, there is access to any corner of the active flavor phase space. }
\label{fig:all}
\end{figure}

Figure~\ref{fig:all} demonstrates our machinery. The left plot shows the four-flavor tetrahedron phase space. Each corner represents pure $\nu_e (1:0:0:0)$, $\numu (0:1:0:0)$, $\nutau (0:0:1:0)$, and $\nu_s (0:0:0:1)$ states, respectively. The mixing is performed under unitarity evolution of $SU(4)$ Haar measure~\cite{Brdar:2016thq}, meaning that all mixings are allowed with a flat distribution. For example, the red volume shows all possible flavor ratios reachable from the pure $\nue$ state at production. As shown in our previous work~\cite{Arguelles:2015dca}, there is a slight overlap between these four volumes. However, they are mostly separated. This means from the measured flavor ratios on Earth, it may be possible to identify the initial flavor state and effects of new physics simultaneously. 

The right plot of Figure~\ref{fig:all} is the observable three-flavor triangle space made from the projection of the four-flavor tetrahedron space (Figure~\ref{fig:all}, left). For this, points in the four-dimensional flavor tetrahedron are projected down onto the active flavor triangle, where $\al_s=0$. The line of projection subtends the sterile corner apex and the point itself. This operation corresponds to renormalizing the flavor ratio $\ofratios$ to $\ofratio$, by rescaling from $\al^{\oplus}_e+\al^{\oplus}_\mu+\al^{\oplus}_\ta+\al^{\oplus}_s=1$ to $\al^{\oplus}_e+\al^{\oplus}_\mu+\al^{\oplus}_\ta=1$. The underlying assumptions are that we do not observe sterile neutrinos and we do not know the absolute astrophysical neutrino flux normalization, but we can measure the three-flavor fraction. Since the sterile component is not observable, we weight the four-flavor tetrahedron points by $(1-\alpha_s)$ when projecting them onto the three-flavor triangle, which is a closer representation of the expected observed flavor triangle. Shown in Figure~\ref{fig:all}, right, this reproduces our previous work in the case of no sterile neutrino states~\cite{Arguelles:2015dca}. For example, when the initial state is $\nue$ only $(1:0:0)$, the available phase space of flavor ratio is confined to the bottom right red area.

The situation is different if there is an initial sterile neutrino state at the production of these astrophysical neutrinos. This is indicated by the purple region, which can distribute across the regions of $\nue$ only $(1:0:0)$, $\numu$ only $(0:1:0)$, and $\nutau$ only $(0:0:1)$. This implies that the BSM mechanism to produce sterile neutrinos at production can violate unitary evolution in three-flavor space, and that can generate new flavor structure on Earth.

This study is different from past studies where non-unitarity is defined as the deviation of the standard oscillation~\cite{Fong:2017gke,Blennow:2016jkn} due to the leakage of probability from active-sterile neutrino mixing. The leakage probability $\mathcal{C}_{\al\be}$~\cite{Fong:2017gke,Blennow:2016jkn} is a matrix of six probability parameters that are explicitly related to appearance and disappearance oscillation experiments. In astrophysical neutrino flavor physics, we are only sensitive to the averaged oscillation from an unknown admixture of neutrino flavors and detailed information of appearance and disappearance are unresolved. In this case, $\ep$ can probe non-unitarity in the leptonic mixing matrix with one parameter. In the future, with the advent of higher statistics and better control of systematics in neutrino telescopes, we will be able to recast this analysis in terms of the different appearance and disappearance scenarios that allow us to rigorously test the elements of $\mathcal{C}_{\alpha\beta}$. 

\section{Flavor ratios with sterile-neutrino mixing but zero initial sterile neutrino fraction ($x=0$)}
\label{sec:mixing1}

In this section, we explore the unitary evolution of neutrino flavors in the four-flavor tetrahedron space. For this, we pick initial flavor ratios without sterile neutrinos, {\it i.e.} $x=0$. This corresponds to the situation where neutrino production mechanisms are bound within the $\nu$SM, but neutrino propagation causes mixing with sterile neutrinos.

\begin{figure*}
\centering
\begin{tabular}{ccc}
\includegraphics[width=0.3\textwidth]{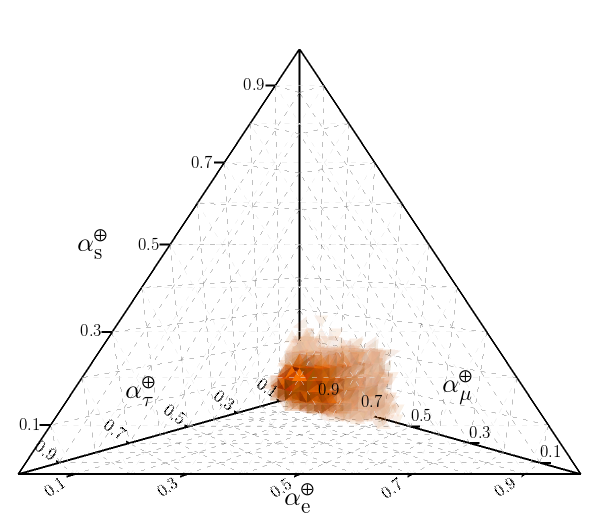} 
& 
\includegraphics[width=0.3\textwidth]{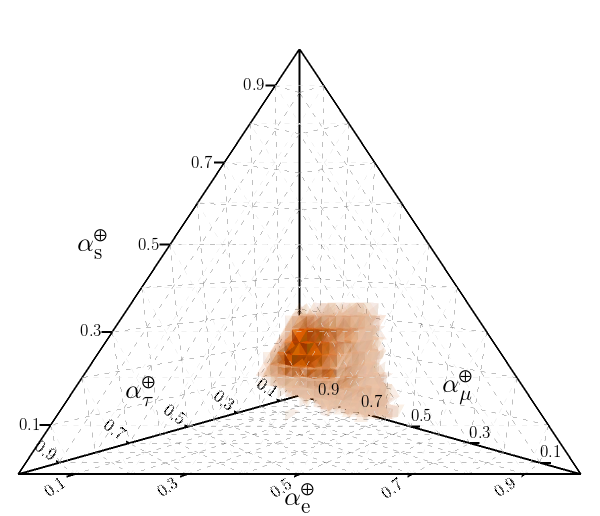} 
&
\includegraphics[width=0.3\textwidth]{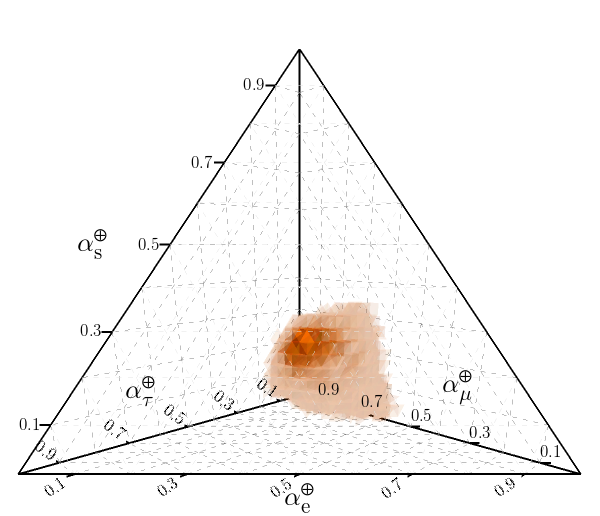}
\\
\includegraphics[width=0.3\textwidth]{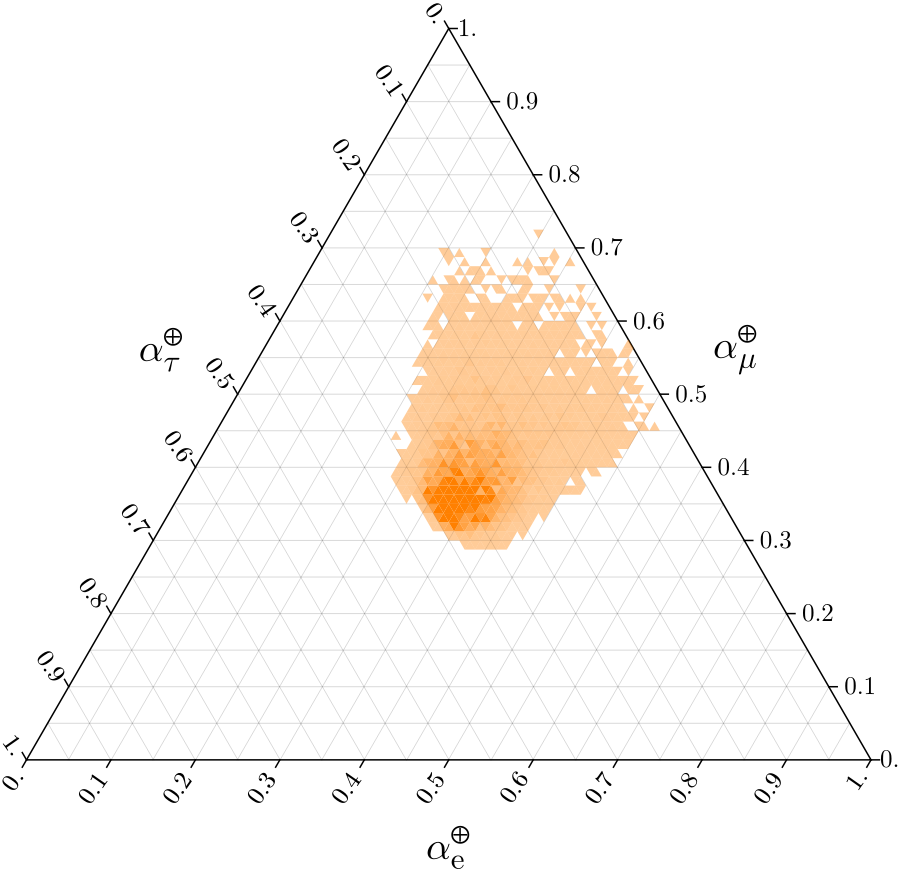}
& 
\includegraphics[width=0.3\textwidth]{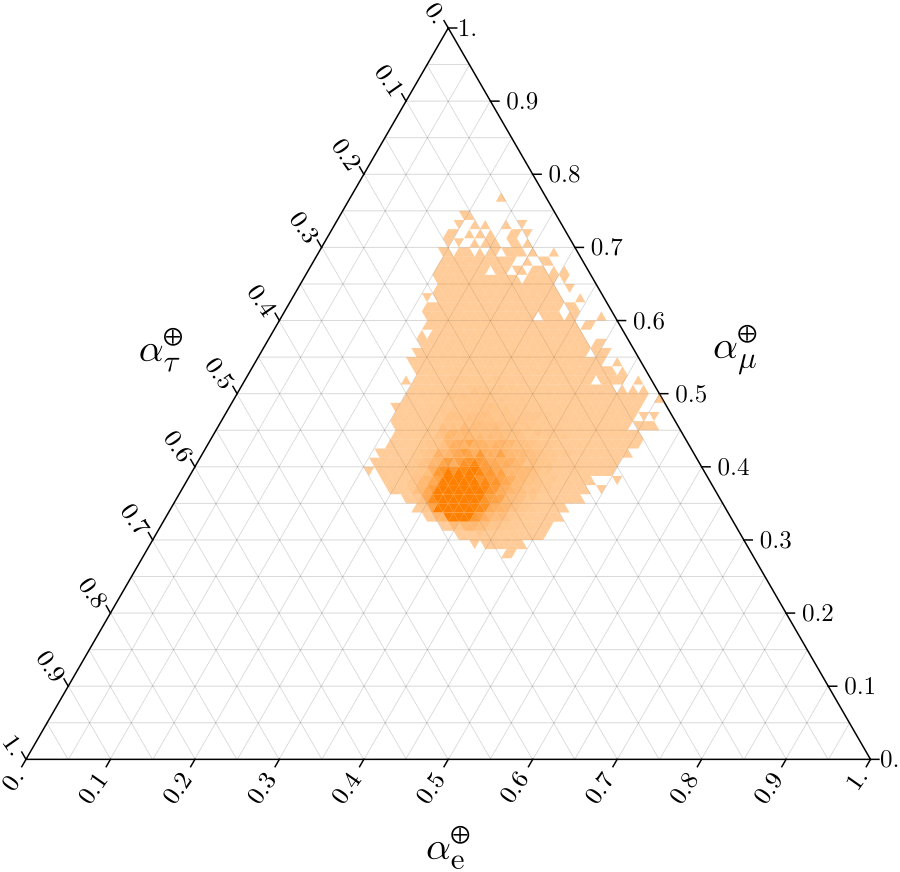}
&
\includegraphics[width=0.3\textwidth]{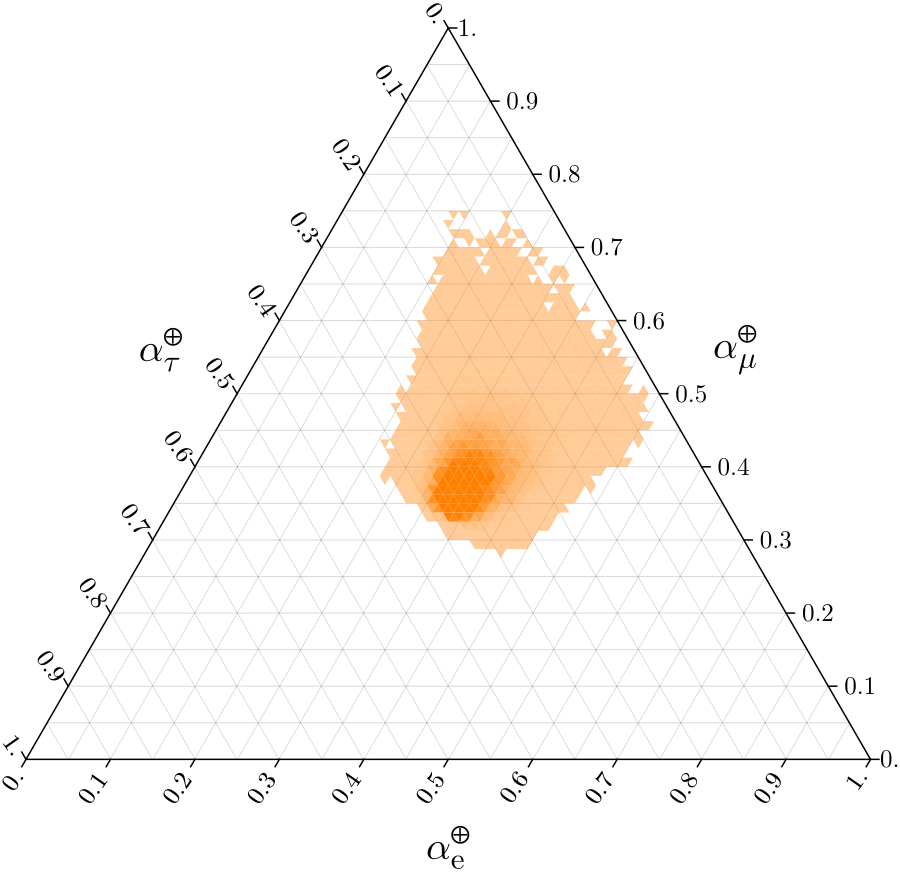}
\end{tabular}
\caption{The observable flavor ratios from given initial source composition ($1:2:0:0$) with $\ep=$ 25\% (left), 50\% (middle), and 75\% (right). The top plots describe flavor ratio distributions in the four-flavor tetrahedron space, and bottom plots are the projections onto three-flavor triangle space.}
\label{fig:1200}
\end{figure*}

Figure~\ref{fig:1200} shows the distributions of possible flavor ratios in the four-flavor tetrahedron space and their projections onto the three-flavor triangle space. In this figure, we use the pion decay scenario ($1:2:0:0$) as the initial flavor ratio. We assume $\ep$ to be fixed at 25\%, 50\%, and 75\%, which describes nonzero mixing with sterile neutrino states through propagation. As shown, for larger values of $\ep$ the phase space tends to disperse upwards in the tetrahedron, {\it i.e.} toward a higher fraction of sterile neutrinos. Upon projection, this causes the possible flavor ratio in the three-flavor triangle state to spread outward.

However, note that this type of mixing does not allow one to move outside of the phase space bounds significantly compared to the unitary evolution for three-active-flavor mixing~\cite{Arguelles:2015dca}. 
The reason that larger $\ep$ values does not modify the distribution extensively here is due to the unitary bound existing in the 4-dimensional flavor space.
In our analysis, we found that increasing $\ep$ causes a redistribution inside the 4-dimensional unitary bound, so it is natural that upon projection, a flavor composition should be confined by the projected unitary bound in the 3-flavor space.
This is due to the fact that, experiments are only sensitive to the active flavors.
The presence of large $\ep$ can also be generalised to the $n$ neutrino types.
Introducing new degrees of freedom results in a phase space, as prescribed by the Haar measure of $SU(n)$, with greater concentration towards centre $(1,\ldots,1)$.
As a result, we expect that projecting to the three active flavors causes our distribution to have more points towards the center.

\begin{figure*}
\centering
\begin{tabular}{ccc}
\includegraphics[width=0.33\textwidth]{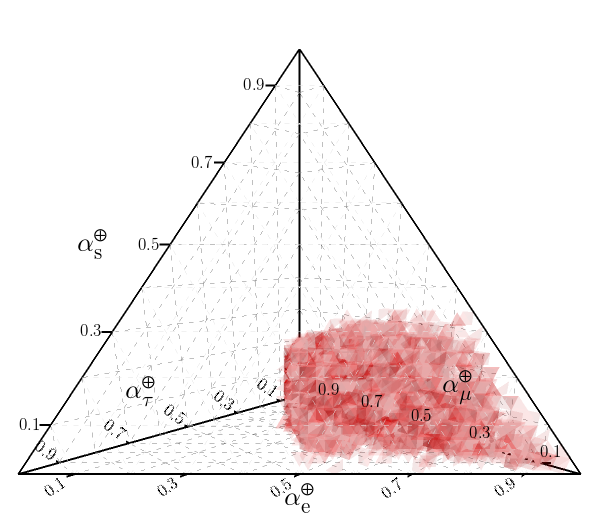} 
& 
\includegraphics[width=0.33\textwidth]{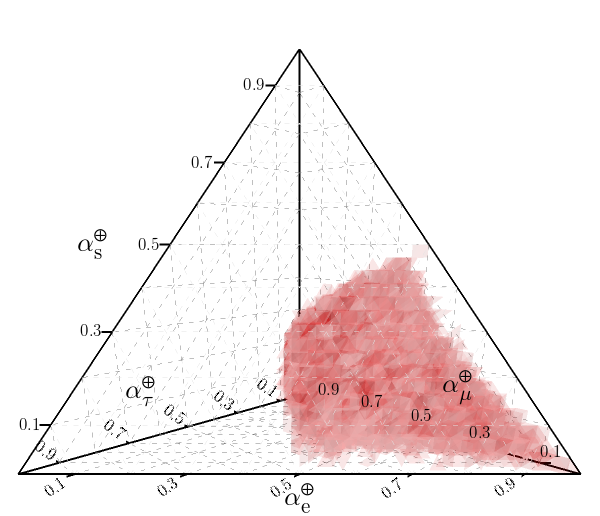} 
&
\includegraphics[width=0.33\textwidth]{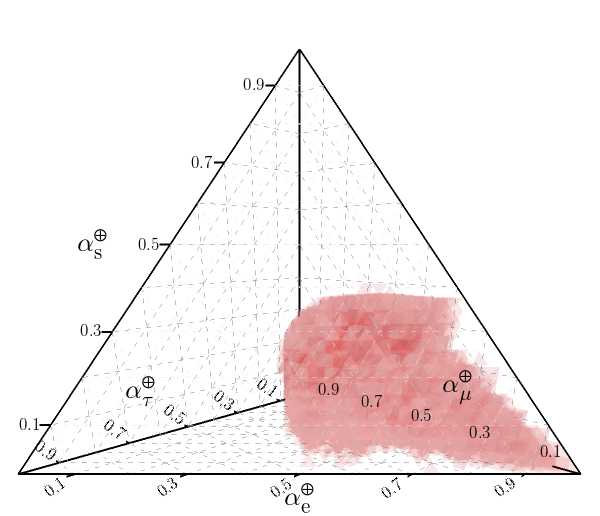}
\\
\includegraphics[width=0.3\textwidth]{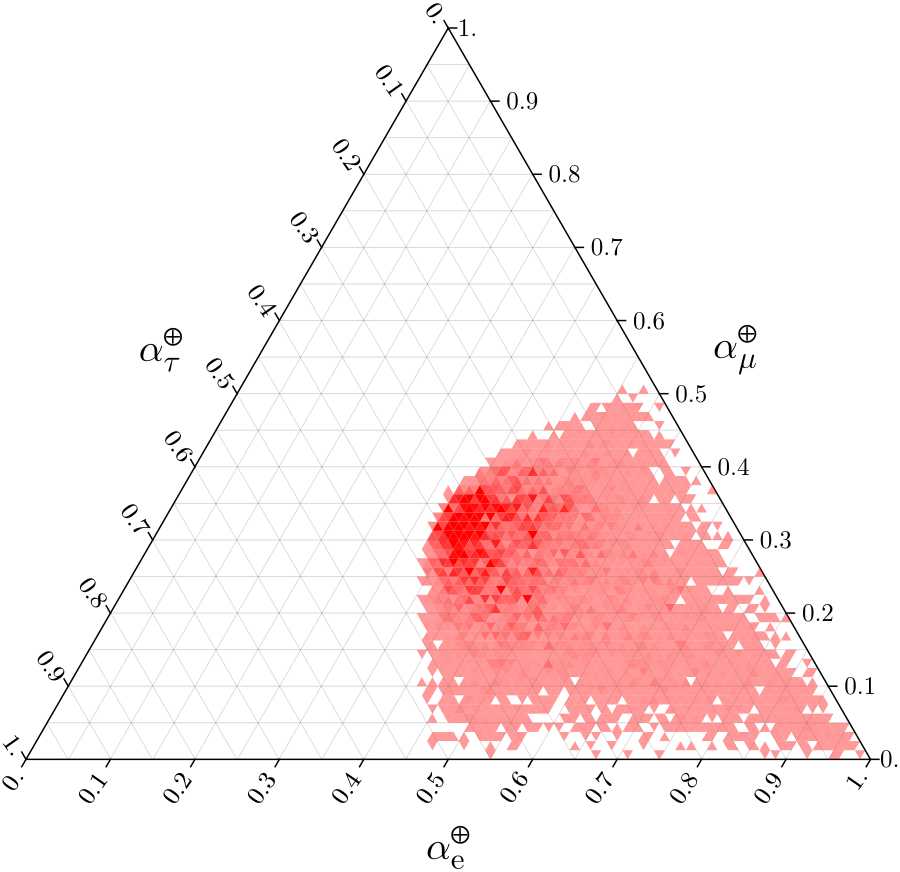}
& 
\includegraphics[width=0.3\textwidth]{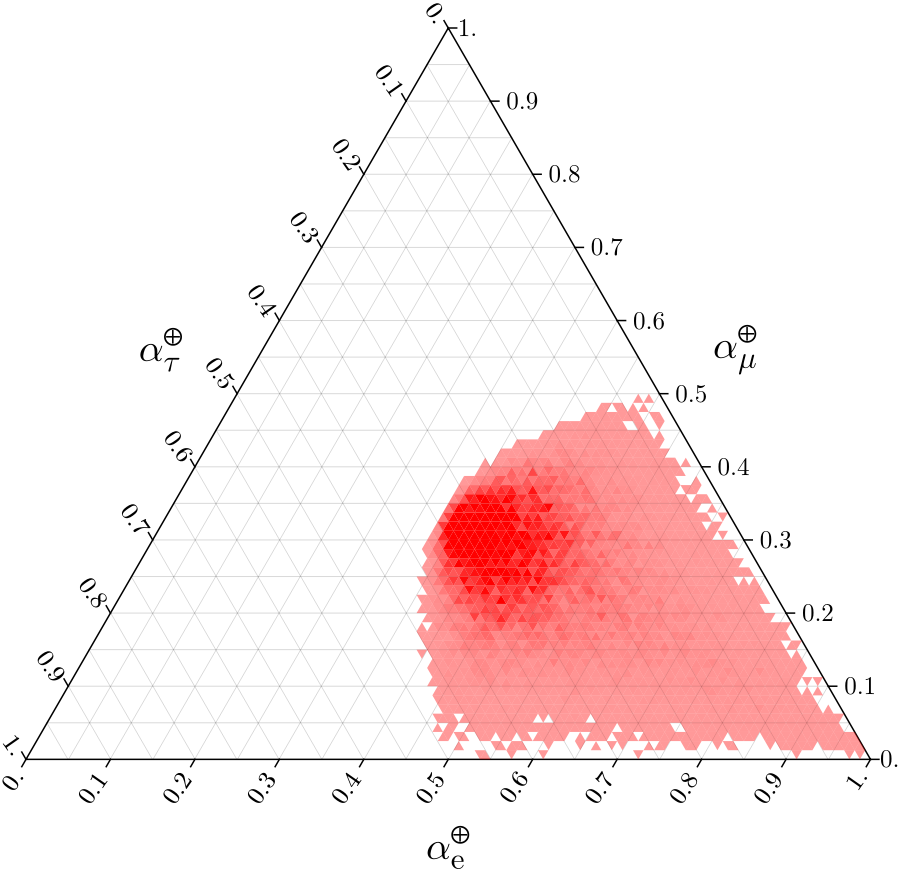}
&
\includegraphics[width=0.3\textwidth]{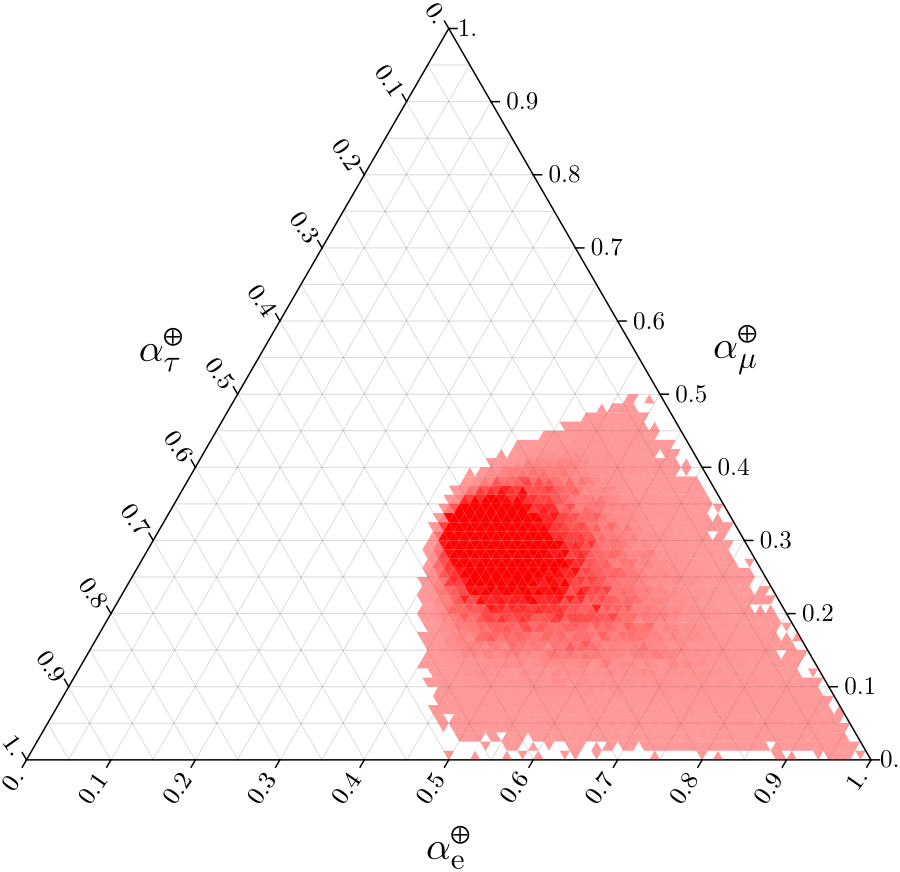} \end{tabular}
\caption{The observable flavor ratios from a given initial source composition ($1:0:0:0$) with $\ep=$25\% (left), 50\% (middle), and 75\% (right). The top plots describe flavor ratio distributions in the four-flavor tetrahedron space, and bottom plots are the projections onto the three-flavor triangle space.}
\label{fig:1000}
\end{figure*}

In summary, nonzero $\ep$  results in a redistribution of density in the flavor ratios. The situation is similar with other initial flavor ratio assumptions, as shown in Figure~\ref{fig:1000}. 

In the literature~\cite{Bustamante:2016ciw}, it has been discussed that in the (3+1) scenario, the distribution of ratios that exist in the three-flavor triangle could breach past the confining regions of available parameter space by mixing. This appears true in the pion production case; however, for the $\nu_e$ dominant scenario, the sterile neutrino mixing is confined to the right bottom corner region. 

\section{Flavor ratios with sterile-neutrino mixing and nonzero initial sterile neutrino fraction ($x\neq 0$)}
\label{sec:mixing2}

\begin{figure}
\centering
\begin{tabular}{ccc}
\includegraphics[width=0.3\textwidth]{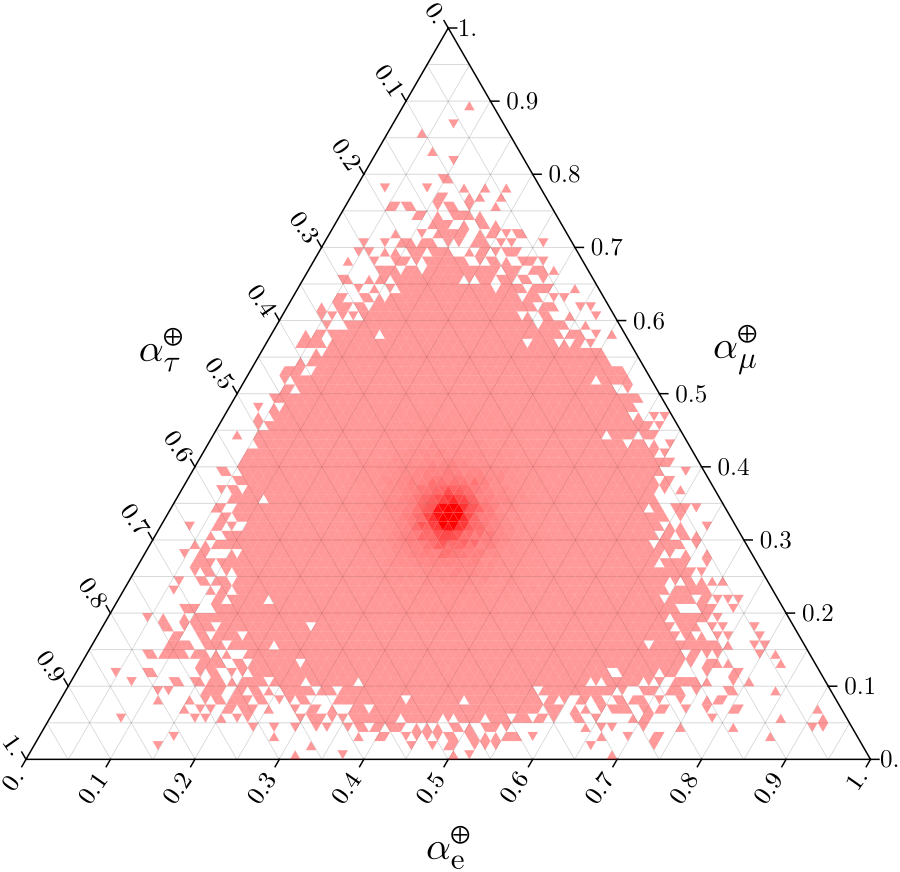}
& 
\includegraphics[width=0.3\textwidth]{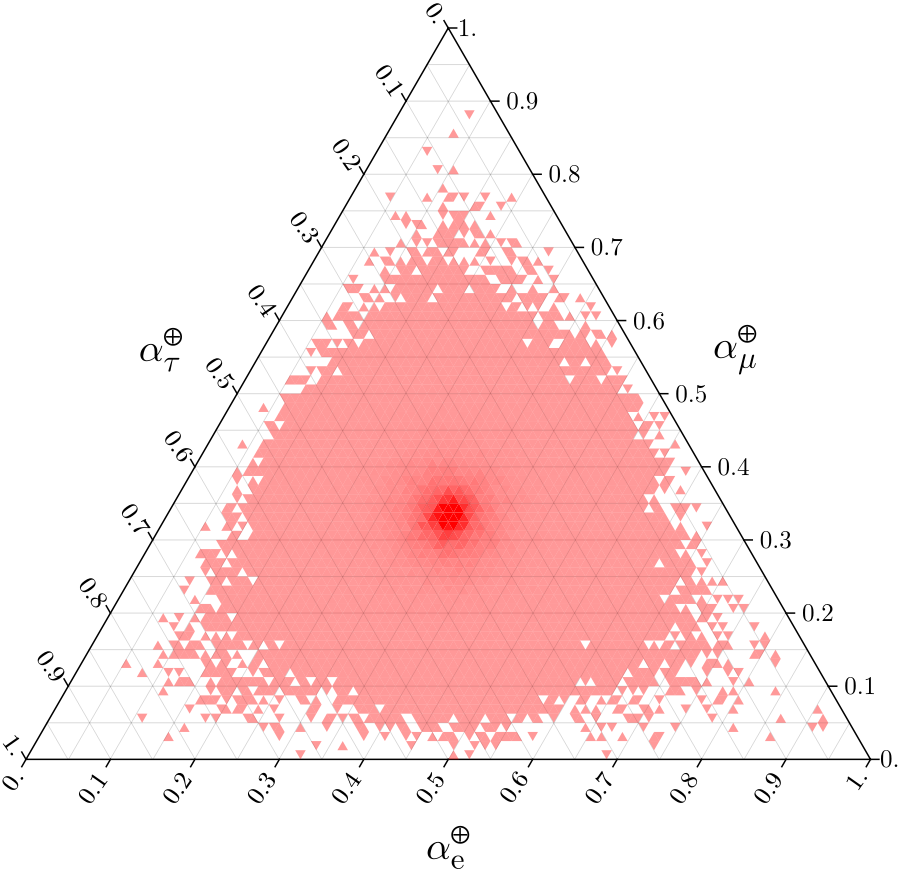}
&
\includegraphics[width=0.3\textwidth]{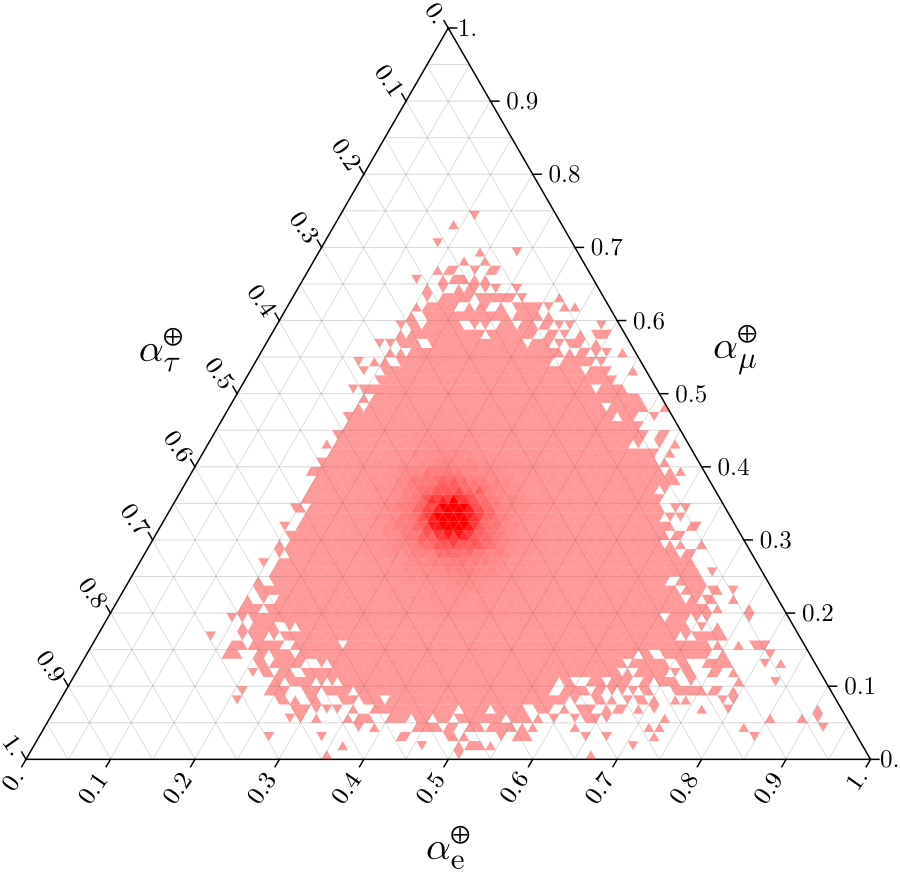} \\
\includegraphics[width=0.3\textwidth]{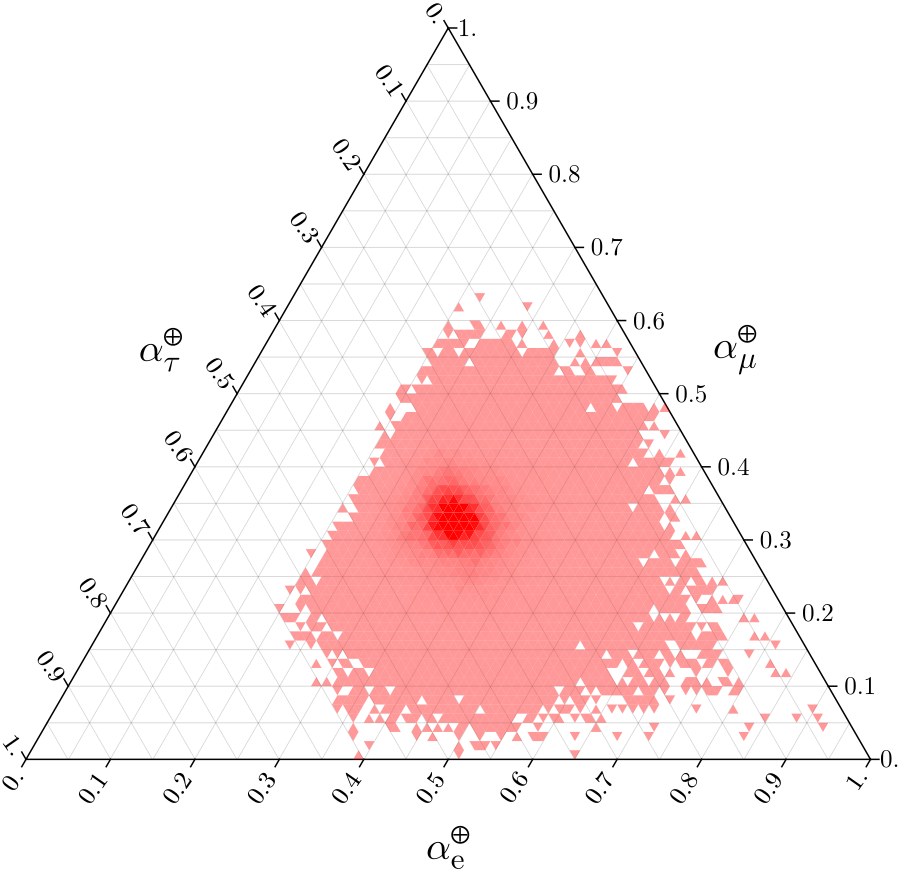}
& 
\includegraphics[width=0.3\textwidth]{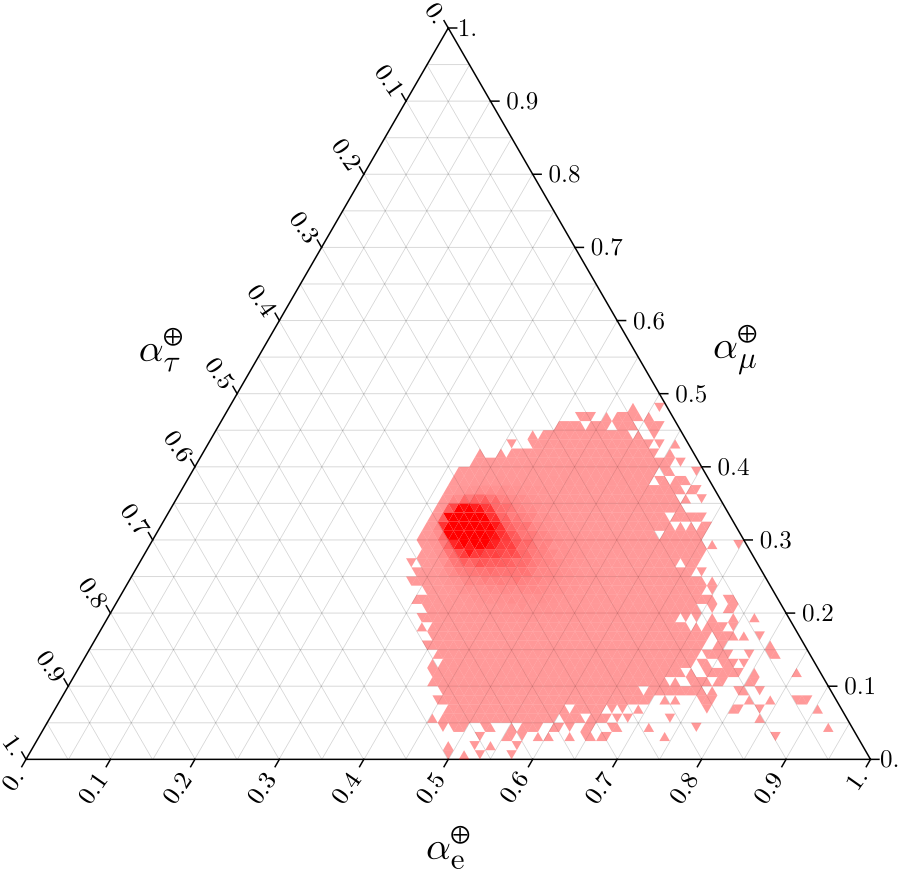}
&
\includegraphics[width=0.3\textwidth]{final_figures/p751000.png}\end{tabular}
\caption{The observable flavor ratios from a given initial source composition ($1:0:0:x$) with $\ep$ set to 75\%. From top left to right bottom, $x=100,50,10,5,1$ and $0$ respectively. All of them are the projections of the four-flavor tetrahedron onto the three-flavor triangle space. If there is a large amount of sterile neutrinos at source, the phase space expands to a wider region. The decrease in the sterile neutrino flavor at the source drives points into the electron neutrino corner of the phase space.}
\label{fig:sterile}
\end{figure}

Figure~\ref{fig:sterile} shows the observable flavor ratios with active-sterile mixing and a nonzero initial sterile neutrino fraction. 
The light sterile neutrino dominated scenario has been proposed, in the literature, as a possible mechanism for neutrino production via the decay of heavy sterile neutrinos that potentially makes up the dark matter content of the Universe~\cite{Brdar:2016thq}.
In this work, we assume that sterile neutrino flavor must have subdominant interactions or noninteractions with standard matter.
This is explicit in how we weight the 4-neutrino points when projecting them to the 3-neutrino flavor space.
In the case that, secret heavy neutrino interactions are assumed, they can produce cascades in neutrino telescopes~\cite{Coloma:2017ppo} and our weighting scheme would need to be adjusted accordingly to account for this degeneracy.
As we see in Figure~\ref{fig:all}, the presence of a nonzero sterile component at production allows non-unitarity within the three-flavor triangle space. In this analysis, we assume the initial flavor ratio is $(1:0:0:x)$ where $x$ varies from 100 to 0. This is such that $x=100$ corresponds to $\sim (0:0:0:1)$, the observable flavor ratio on Earth (projection of the four-flavor tetrahedron onto the three-flavor triangle space) can be anything, although the phase space density is focused around the center. As $x$ decreases, and eventually becomes 0, the allowed phase space is identical to what one would expect from the three-active-flavor case. 

\section{Flavor ratios with experimental constraints}
\label{sec:mixing3}

In this section, we discuss the expected flavor ratio on Earth. First, we start the discussion in the case of unitarity for the three-flavor paradigm. For this, we sample the mixing matrix elements within the $3\si$ C.L. range from the global fit of the PMNS matrix (Eq.~\eqref{eq:unitarity})~\cite{Esteban:2018azc}. The nine conditions are applied independently without assuming correlations. This may overestimate the area but the expected effect is small. Then Eq.~\eqref{eq:prob} is used to compute the mixing probability. Since the flavor ratio of astrophysical neutrinos at the source is not known, we generalize this by assuming $\ifratios=(y:1-y:0:0)$. Namely, it includes all possible combinations of $\nue$ and $\numu$ but not $\nutau$ and $\nu_s$. The blue region in Figure~\ref{fig:moneyplot} shows the result. This is consistent with previous work~\cite{Arguelles:2015dca,Bustamante:2015waa}.

Now we relax unitarity, by applying the nine constraints on the PMNS matrix elements~\cite{Parke:2015goa} (Eq.~\eqref{eq:nonunitarity}). We use them as independent constraints and follow the same procedure described above to make the orange region in Figure~\ref{fig:moneyplot}. In this figure, we also plot the measured flavor ratio from IceCube~\cite{Aartsen:2015knd}. We note that the relaxation of the unitarity hypothesis in the active-neutrino sector expands the allowed region and calculate that the maximum $\ep$ allowed in this scenario (assuming non-unitarity) is approximately $\ep \lesssim 0.51$. Any higher value than this for non-unitarity violates the global constraints at the $3\si$ level. One key topological effect is the distribution of the allowed region is still forbidden to reach the $\phi_\tau$ corner of phase space. This is a known feature within the three-flavor active-neutrino mixing scenario where initial production of $\nutau$ is required to cover this region. As we saw in the case of Figure~\ref{fig:sterile}, it is also possible to access this region if the sterile neutrino fraction produced at the source is significantly larger than the active components. Notice that both the blue and orange regions have very similar profiles, though due to the relaxation of unitarity, mainly in the $U_{\ta 1}$, $U_{\ta 2}$, and $U_{\ta 3}$, the region has enlarged towards points with larger contribution of muon and smaller tau neutrino fluxes at Earth. This is expected due to the greater range these parameters have in the PMNS matrix and the nonexistence of an initial $\nutau$ state. 

\begin{figure}
\centering
\includegraphics[width=0.4\textwidth]{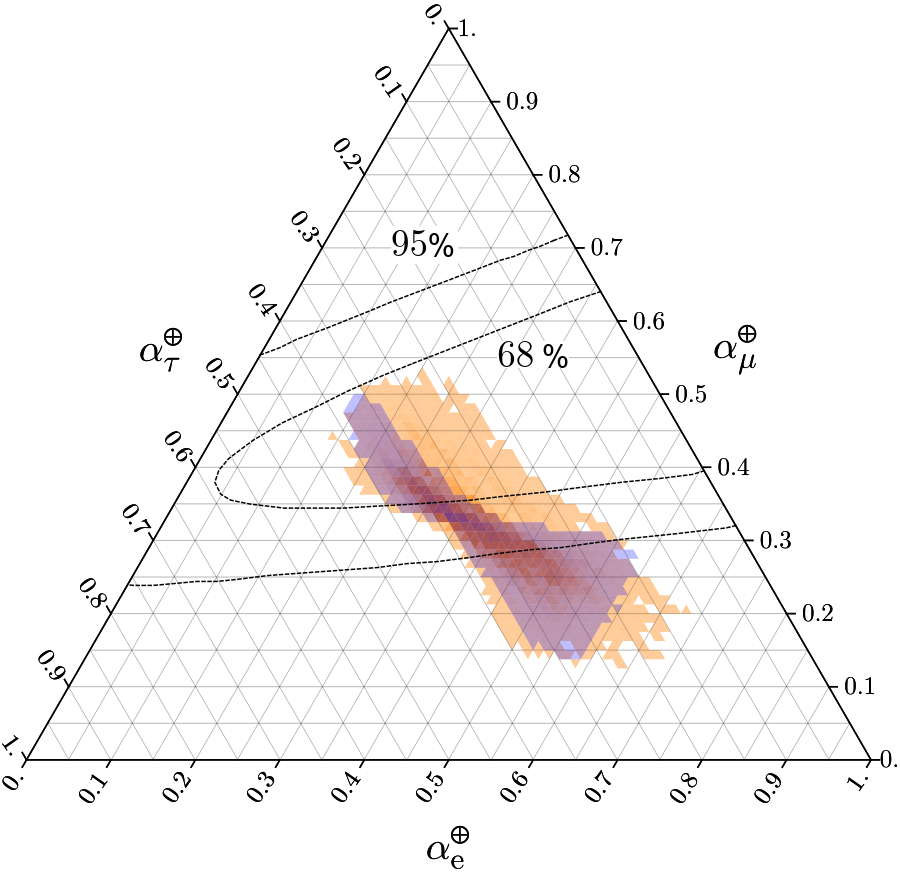}
\caption{The figure shows the distribution of expected astrophysical neutrino flavor ratios with (blue) and without (orange) unitarity of the PMNS matrix. Matrix elements are sampled within the $3\si$ range from the global fit with~\cite{Esteban:2018azc} and without~\cite{Parke:2015goa} unitarity. We use $\ifratios=(y:1-y:0:0)$ ($0<y<1$) for the source flavor ratio. These have been superimposed on the measured flavor ratio from IceCube~\cite{Aartsen:2015knd} showing 68\% and 95\% contours.}
\label{fig:moneyplot}
\end{figure}

\section{Discussion and conclusions}
\label{sec:conclusions}

In this work, we have presented a series of studies of astrophysical neutrinos assuming the presence of sterile neutrino states through flavor ratios observable on Earth. We performed a study of the implications of the presence of sterile neutrinos on producing non-unitarity in the astrophysical flavor content. We compute the unitary evolution of four neutrino states within a four-flavor tetrahedron space. Then, a projection of this onto the three-flavor triangle space describes observable states. 
\begin{itemize}
\item[1] This shows that allowing mixing with sterile neutrinos would redistribute the possible flavor ratio mostly within the confined phase space not far away from the one defined in three-flavor unitarity. This indicates that the effect of active-sterile neutrino mixing is not easy to study from the measured flavor ratio on Earth without assuming a significant amount of sterile neutrino production at the source. 
\item[2] The significant unitarity violation can be introduced when we include a nonzero sterile neutrino at the source. Thus, astrophysical neutrino flux models including sterile neutrino states may be interesting to investigate in the future.
\item[3] Finally, we illustrate the possible flavor ratio given global data relaxing unitarity constraints. This shows that the allowed phase space is expanded in the direction of less $\nutau$. 
\end{itemize}

The method we used is suitable for current and future searches of sterile neutrino states through astrophysical neutrino flavors. In particular, astrophysical neutrino flavors can investigate sterile neutrinos with any masses, including extremely small mass splitting. Interestingly, such sterile neutrinos could be potential candidates for dark matter in the universe. Future large arrays, such as KM3NeT~\cite{Adrian-Martinez:2016fdl} and IceCube-Gen2~\cite{Aartsen:2014njl,TheIceCube-Gen2:2016cap}, can further investigate these areas.

\acknowledgments
We thank to Jean DeMerit for proofreading a draft. CA is supported by NSF grant PHY-1912764. KF acknowledge financial support from the NExT/SEPnet Institute. TK and SM acknowledge financial support from the Science and Technology Facilities Council (STFC), United Kingdom (UK). RK acknowledges financial support from the Wisconsin IceCube Particle Astrophysics Center (WIPAC). This project is supported by the international exchanges grant from the Royal Society, UK. JS acknowledges financial support from EU Networks FP10ITN ELUSIVES (H2020-MSCA-ITN-2015-674896) and INVISIBLES-PLUS (H2020-MSCA-RISE2015-690575), by the MINECO grant FPA2016-76005-C2-1-P and by the Maria de Maeztu grant MDM-2014-0367 of ICCUB.

\bibliographystyle{JHEP}
\bibliography{nonunitarity}

\appendix

\section{Generating matrix according to the Haar measure for $n$ neutrino types}

In this section we describe the algorithm used in this work to construct the sets of points shown in our figures, when sampling according to the Haar measure and when imposing further constraints.

\begin{enumerate}
    \item Generate a unitary matrix under the unitary Haar measure - here we sample $n^2$ pairs of independent and identically distributed (i.i.d.) random variables (r.v.s) $x_k,y_k$ for $k=1,\ldots,n\times n$, according to the standard normal distribution:
    \begin{equation}
        x_k,y_k \sim \mathcal{N}(\mu=0,\sigma^2=1).
    \end{equation}
    Next we define for each $k=1,\ldots,n^2$ a standard complex normal r.v., $z_k$ via the sum
    \begin{equation}
        z_k = x_k + i y_k \in \mathbb{C},
    \end{equation}
    We then construct a $n\times n$ unitary matrix using these $n^2$ variables, namely
    \begin{equation}
        {\mathbf{Z}} = \left(\begin{matrix}
        z_{1\times 1} &\ldots &z_{1\times n}\\
        \vdots & \ddots & \vdots \\
        z_{n\times 1} & \ldots & z_{n\times n}
        \end{matrix}\right).
    \end{equation}
    
Since ${\mathbf{Z}}$ is of full rank and it's entries are i.i.d. as standard complex r.v.s, it forms part of the Ginibre ensemble, \textit{i.e.} the set of $n\times n$ matrices with gaussian i.i.d r.v.s whose eigenvalues tend to be uniformaly distributed over the unit disk as $n\rightarrow \infty$. Hence, we can orthonormalize its columns using the Gram-Schmidt algorithm, and produce a uniformly sampled unitary matrix:
\begin{equation}
    \mathbf{Z} = \mathbf{U}\cdot \mathbf{Q},
\end{equation}
where $\mathbf{U}$ is our uniformly sampled unitary matrix given by
\begin{equation}
    \mathbf{U} = \left(
    \begin{matrix}
    u_{1\times1}& \ldots & u_{1\times n} \\
    \vdots & \ddots & \vdots \\
    u_{n\times 1} & \ldots & u_{n\times n},
    \end{matrix}
    \right),
\end{equation}
and $\mathbf{Q}$ is a upper triangular  invertible matrix.
\item Computation of $\epsilon$: To compute $\epsilon$, we take the principal submatrix of $\mathbf{U}$, denoted here as $U$, which is formed by deleting the last row and column of the unitary matrix $\mathbf{U}$: 
\begin{equation}
   U = \left(
    \begin{matrix}
    u_{1\times1}& \ldots & u_{1\times (n-1)} \\
    \vdots & \ddots & \vdots \\
    u_{(n-1)\times 1} & \ldots & u_{(n-1)\times (n-1)}
    \end{matrix}
    \right).
\end{equation}
We then compute
\begin{equation}
     \epsilon =||U\cdot U^\dagger - \mathds{1}||_2:= \max_{|\mathbf{v}|_{2}\neq 0} \frac{|(U\cdot U^\dagger - \mathds{1})\mathbf{v}|_2}{|\mathbf{v}|_2},
\end{equation}
where $\mathbf{v}=(v_1,\ldots ,v_n)\in \mathbb{C}$ and $|\mathbf{v}|_2 := \sqrt{\sum_{k=1}^{n}v_k^2}$.
Note that $\epsilon$ is also the square root of the maximum eigenvalue for the square matrix $(U\cdot U^\dagger - \mathds{1})^{\dagger}(U\cdot U^\dagger - \mathds{1})$ and can be computed this way also. We then collect the matrices according to their $\epsilon$ value.
\item Calculating probability and flavour ratio: we compute the probability according to Eq.~\eqref{eq:prob}. Given some initial flavor ratio, we then calculate the expectation for the terrestial flavor ratio as in Eq.~\eqref{eq:earthratio}.
\item Projecting the point: Given some flavor ratio, $\ofratios$, we can project the point onto the tetrahedron space using the following projection map:
\begin{equation}
    \pi_{tetra}(\phi_e^\oplus,\phi_\mu^\oplus,\phi_\tau^\oplus,\phi_s^\oplus) =\left(\sum_{\alpha=e,\mu,\tau,s}\phi_\alpha^\oplus\right)^{-1} \left(\begin{matrix}\dfrac{1}{2}(2\phi_e^\oplus + \phi_\mu^\oplus + \phi_s^\oplus) \\
    \dfrac{\sqrt{3}}{6}(3\phi_\mu^\oplus + \phi_s^\oplus)\\
    \dfrac{\sqrt{6}}{3}\phi_s^\oplus
    \end{matrix}\right).
\end{equation}
Likewise, projecting this flavour ratio to the active-flavor triangle, we can use the projection map
\begin{equation}
    \pi_{tri}(\phi_e^\oplus,\phi_\mu^\oplus,\phi_\tau^\oplus,\phi_s^\oplus) = \left(\sum_{\alpha=e,\mu,\tau}\phi_\alpha^\oplus\right)^{-1}\left(\begin{matrix}
    \dfrac{1}{2}(2\phi_\mu^\oplus + \phi_\tau^\oplus)\\
    \dfrac{\sqrt{3}}{2}\phi_\tau^\oplus
    \end{matrix}\right).
\end{equation}
\item Binning: To visualise the distribution of the points, we bin in equivolumetric tetrahedral elements of the flavor tetrahedron, or equilateral triangles in the projected active flavor triangle. In turns out that a tetrahedron can be decomposed into two types of tetrahedral components with equal volume but different topology, and hence we adopt this polyhedral mesh for the tetrahedron.

Next we compute the number of points in each bin and weight the bin opacity by the number of points in each bin. The $j$-th bin opacity $o_j$ is defined by
\begin{equation}
    o_j=\frac{n_j}{\sum_k n_k},
\end{equation}
where $n_j$ is the number of points in the $j$-th bin.
For all $j$ we see $o_j \in [i_{\min},i_{\max}]$ where $i_{\min}=0\%$ and $i_{\max}=100\%$.

Many bins have almost zero opacity, and in the central region of the tetrahedron and triangle phase spaces, the opacity is close to one. To improve visualisation of these extreme regions, we scale all the opacities to fit in the range $i'_{\min}=10\%$ to $i'_{\max}=90\%$ by the formula
\begin{equation}
    o_j \mapsto o_j'=\frac{o_j - i_{\min}}{i_{\max}-i_{\min}} \times (i'_{\max} - i'_{\min}) + i'_{\min},
\end{equation}
and hence set the opacity of each bin to be $o_j'$.
\end{enumerate}
To then further constrain the point to satisfy a given non-unitarity amount, namely to satisfy Eq.~\eqref{eq:nonunitarity}, in between steps 2 and 3, we use the following procedure:
\begin{enumerate}
    \item We first consider the $\chi^2$-distributions for the matrix parameter fit using the constraints from Ref.~\ref{eq:nonunitarity}.
    \item We compute the distribution $\exp(-\chi^2)$ for each of the matrix parameters, and perform rejection sampling by constructing a box around each exponentiated distribution. Afterwards we accept matrices satisfying the rejection sampling range and plot these in our distribution.
    \end{enumerate}
\end{document}